\documentstyle
[12pt]
{article}
\begin{document}
\newcommand{\ed}{\end{document}}
\newcommand{\pr}{\prime}
\newcommand{\ppr}{\prime\prime}
\newcommand{\cE}{{\cal E}}
\newcommand{\vphi}{{\varphi}}
\newcommand{\oO}{O(k^{-1})}
\newcommand{\be}{\begin{equation}}
\newcommand{\ee}{\end{equation}}
\newcommand{\barr}{\begin{array}}
\newcommand{\earr}{\end{array}}
\newcommand{\bea}{\begin{eqnarray}}
\newcommand{\eea}{\end{eqnarray}}
\newcommand{\bean}{\begin{eqnarray*}}
\newcommand{\eean}{\end{eqnarray*}}
\newcommand{\pa}{\partial}
\newcommand{\enm}{energy-momentum tensor}

\title{Generalized abelian coset conformal field theories.}
\author{A.V.Bratchikov
\thanks{bratchikov@kubstu.ru}\\ Kuban State Technological University,\\
2 Moskovskaya Street, Krasnodar, 350072, Russia.} \date{December 1999}
\maketitle \begin{abstract}
The reductions of conformal field theories which lead to
generalized abelian cosets are studied.
Primary fields and correlation functions of
arbitrary abelian coset conformal field theory are explicitly
expressed in terms of those of the original theory.
The coset theory has global abelian symmetry.
\end{abstract}

{\bf PACS} 11.25.Hf.

{\bf Keywords:} conformal field theory, coset models,extended symmetry
algebras.

In this paper we study the conformal field theory based on the
generalized coset construction \cite {BH,GKO}
\bea \label{K}
K(m)=L(m)-L^{u}(m),\quad m\in Z,
\eea
where $L(m)$ , $L^{u}(m)$ are Virasoro generators and
$L^{u}(m)$ are quadratic in affine $\hat u(1)^d$ currents. (For a
review see \cite {BS,HKOC}).

When $L(m)$ are given  by Sugawara construction
for the affine Lie algebra $\hat g$ we have the $g/u(1)^d, 1\le  d
\le rank \,g,$ coset.
The $su(2)/u(1)$ theory is equivalent to the
$Z_k$ invariant parafermion theory whose studies were initiated by
Fateev and Zamolodchikov \cite {FZ}.
Various aspects of $g/u(1)^d$ coset theories for arbitrary simple $g$
were investigated in Refs.[6-10].

There are many other conformal models which have $\hat u(1)^d$
affine subalgebras and can be reduced to coset (\ref {K}).The class of
such models includes $N=2$ superconformal algebras \cite {A},
$N$-extended superconformal algebras \cite {B,K} , non-semi-simply
affine-Sugawara constructions \cite {ORS,FS} and other.
Classical algebras with Virasoro generators (\ref {K})
were studied in Ref.\cite {D}.Some currents of the reduced $N=2$ and
$W^{sl(3)}_{2,1}$ algebras were obtained in Ref.\cite {BEHHH}.
As well as the original models the reduced ones can be used as
building blocks in string theories.
Therefore it is important to investigate their properties.

The object of this paper is to construct a class of primary fields of
(\ref {K}) and corresponding correlation functions in terms of the
original theory based on $L(m).$ To find the
primary fields we use their transformation properties under coset
conformal algebra. Solving corresponding  equations we
express primary fields of (\ref {K}) in terms of the original
primary fields and
$\hat u(1)^d$ currents.
This enables us to express
correlation functions in terms of the original
ones.All the coset fields under consideration commute with the
Heisenberg algebra generated by non-zero modes of  $ \hat u(1)^d$
and form (non-trivial) representations of $u(1)^d.$ As a concequence
of this construction we obtain a new realization of
$Z_k$ invariant parafermion currents.

Let $\Omega$ be the (super)algebra which has the Virasoro and
$\hat u(1)^d$ subalgebras
\bea  \label {V}
[L(m),L(n)]&=&(m-n)L(m+n)+c\left[\frac 1 {12}
(m^3-m)\right]\delta_{m,-n},
\eea
\bean
[J^i(m),J^j(n)]&=&km\delta^{ij}\delta_{m,-n},
\eean
\bean
[L(m),J^i(n)]=-nJ^i(m+n),
\eean
where $m,n\in Z$ and $1\le i,j \le d;$ $c$ and  $k$ are the central
charges.

Let $G_{h}(w)$ be the primary field of $\Omega$ which satisfies the
equations
\bean \label{prV}
[L(m),G_{h}(w)]=w^{m+1}\partial_w G_{h}(w) +h (m+1)w^m
G_{h}(w),
\eean
\bean \label{prA}
[J^i(m),G_{h}(w)]= w^{m}t^i G_{h}(w).
\eean
Here $h$ is the conformal dimension of  $G_{h}(w)$
and $t^i$ is the (reducible) representation of the
generators of $u(1)^d$ for the field $G_h(w).$

$G_h(w)$ can be decomposed in
the set of one-dimensional irreducible representations of $\hat
u(1)^d.$ Let ${G^{\mu}_h(w)}={P_\mu G_h(w)}$ be the irreducible
component of $G_h(w)$ which satisfies the equation \bean \label{prA}
[J^i(m),G_{h}^{\mu}(w)]=\mu^i w^{m}G_{h}^{\mu}(w).
\eean
Here  $P_\mu$ is the projector and $\mu=(\mu^i)$ is the weight.
As well as  $G_h(w)$ the field $G^{\mu}_h(w)$ is primary
with respect to  Virasoro algebra (\ref{V})
\bea \label{L}
[L(m),G_{h}^{\mu}(w)]=w^{m+1}\partial_wG_{h}^{\mu}(w) +h (m+1)w^m
G_{h}^{\mu}(w).
\eea
Correlation functions of these fields can be computed
using correlation functions of primary fields of
$\Omega$
\bean \label{cb} <
G^{\mu_1}_{h_1}(w_1)\ldots G^{\mu_n}_{h_n}(w_n)>= \prod_{i=1}^n
{P_{\mu_i}} < G_{h_1}(w_1)\ldots G_{h_n}(w_n)> \eean
 We shall use the following properties of the vacuum state $\vert 0>$
\bea
\label {vac} <0\vert J^i(m<0)= J^i(m \ge 0) \vert 0> =0.
\eea

The operator $L^{u}(m)$ is given by
 \bean
\label{}
{L^{u}(m)}=\frac 1 {2k} \sum_{i=1}^d {:J^i(m-n)J^i(n):}  .
\eean
The normal-ordering symbol :\,: means that negative modes of
the currents are on the left.  The operators $L^{u}(m)$ satisfy
Virasoro algebra (\ref {V}) with the central charge $c=d$ and \bean
[L^{u}(m),J^i(n)]=-nJ^i(m+n). \eean

We shall use the relation
\bea \label{ue}
[L^u(m),G^\mu_h (w)]= \frac 1
{k} w^{m+1}:\mu \cdot J(w)G^\mu_h(w):+\frac {\mu^2} {2k} (m+1)w^m
G^\mu_h(w) , \eea where \bean :J^i(w)G^\mu_h(w):= \sum_{m<0}{
J^i(m) w^{-m-1}G^\mu_h(w) +G^\mu_h(w)\sum_{m\ge 0} J^i(m) w^{-m-1} }.
\eean

The coset generators  $K(m)$ (\ref {K})
satisfy Virasoro algebra (\ref {V}) with the central
charge $c-d$ and
\bea \label{o}
[J^i(m),K(n)]=0.
\eea
This equation suggests
that all the fields of the coset model commute with $J^i(m).$ However,
the primary fields $\tilde G^\mu_h(w)$ which we
present in this paper commute only
with the Heisenberg
algebra generated by $J^i(m),\,m\ne 0,$ \bea \label{e2} [J^i(m),\tilde
G^\mu_h(w)]=0
\eea
and transform  under $u(1)^d$ as
\bea \label{e3} [J^i(0),\tilde
G^\mu_h(w)]=\mu^i\tilde G^\mu_h(w),
\eea
Eqs.(\ref {o}) and (\ref {e3}) express the fact that the coset
theory has global abelian symmetry.

By the definition the coset primary field $\tilde G^\mu_h(w)$
satisfies the equation
\bea \label{e1} [K(m),\tilde
G^\mu_h(w)]=w^{m+1}\partial_w\tilde G^\mu_h(w) +\tilde h
(m+1)w^m \tilde G^\mu_h(w), \eea
where $\tilde h$ is the conformal dimension of $\tilde
G^\mu_h(w)$ .

Here we give a solution of this equation
\bea \label{sol}
\tilde G^\mu_h(w)&=& U_<^\mu(w)G^\mu_h(w)U^\mu(w), \nonumber \\
\tilde h& =& h -\frac {\mu^2 } {2k} ,
\eea
where
\bean \label{}
U_<^\mu(w)&=&e^{-{\frac 1 k \mu \cdot Q_<(w)}},
\qquad  Q_<^i (w)=-\sum_{m<0}{\frac 1
m J^i(m) w^{-m}},  \nonumber \\
U^\mu(w)&=&U_0^\mu(w)U_>^\mu(w),\qquad  U_0^\mu(w)=w^{-\frac 1  k
\mu \cdot J(0)}, \nonumber \\
U_>^\mu(w)&=&e^{- \frac 1 k \mu
\cdot Q_>(w)}, \qquad Q_>^i(w)=-\sum_{m>0}{\frac 1 m J^i(m)
w^{-m}}.
\eean

To check (\ref {e1}) we have the following computations
$$
\begin {array} {lcl}
[K(m),\tilde G^\mu_h(w)]&=&U^\mu_<(w)\,[K(m),G^\mu_h(w)]\,U^\mu(w)
\\
&=& w^{m+1}U^\mu_<(w)\left(\partial_wG^\mu_h(w)-
\frac 1 k :\mu \cdot J(w)G^\mu_h(w):\right)U^\mu \\
&&+\tilde h (m+1)w^m \tilde G^\mu_h(w)
\\
&=& w^{m+1}\partial_w\tilde G^\mu_h  (w)
+\tilde h (m+1)w^m \tilde G^\mu_h(w).
\end{array}
$$
Here we used eqs. (\ref {o}),(\ref {ue}) and at the last step  the
relations
\bean \pa_w  U^\mu_<(w)&=&-\frac 1 k U^\mu_<(w)\sum_{m<0}{\mu
\cdot J(m)} w^{-m-1}, \nonumber \\ \pa_w  U^\mu(w)&=&-\frac 1 k
U^\mu(w)\sum_{m \ge 0}{\mu \cdot J(m)} w^{-m-1} .  \eean
 It is easy to
check that  $\tilde G^\mu_h(w)$  also satisfies eqs. (\ref {e2})
and (\ref {e3}).

The original primary field $G^\mu_h(w)$ can be expressed as follows
\bean   \label{}
G^\mu_h(w)= \left(U_<^\mu(w)\right)^{-1} \tilde G^\mu_h(w)
\left(U^\mu(w)\right)^{-1} .
\eean

Computations show that
\bea        \label{use}
 U_0^\mu(z) G^\nu_h(w)& =&G^\nu_h(w)U_0^\mu(z)z^{-\frac {\mu\nu} k}
,\nonumber \\
G^\mu_h(z)U^\nu_<(w)&=&
U^\nu_<(w) G^\mu_h(z)\left(1- \frac w z\right)^{-\frac {\mu\nu} k}.
\eea
From this, eqs.(\ref {vac}) and (\ref {e2})
 it follows that the $n-$ point coset correlation
function can be written as \bean \label{} <\tilde
G^{\mu_1}_{h_1}(w_1)\ldots\tilde
G^{\mu_n}_{h_n}(w_n)>= <G^{\mu_1}_{h_1}(w_1)\ldots G^{\mu_n}_{h_n}(w_n)>
\prod_{i<j}{(w_i-w_j)^{-\frac {\mu_i
\mu_j} k  }} .
\eean
In the case of $su(2)/u(1)$ this relation was obtained in Ref.
\cite{FZ}.

Using eqs.(\ref {e2}) and  (\ref {use}) one can find the
operator product expansion of two
coset primary fields \bea
\label{OPE} \tilde G^{\mu_1}_{h_1}(z)\tilde
G^{\mu_2}_{h_2}(w)= (z-w)^{-\frac {\mu_1\mu_2} k} U^{\mu_1}_<(z)
U^{\mu_2}_<(w)G^{\mu_1}_{h_1}(z)G^{\mu_2}_{h_2}(w)
U^{\mu_1}(z)U^{\mu_2}(w) .
\eea

As an example let us consider the $su(2)/u(1)$ coset.The $\hat {su}(2)$
algebra is generated by the currents $E^+(w),E^-(w)$ and $J(w)$ which
satisfy the operator product expansions \bea \label{km}
E^+(z)E^+(w)&=&reg.,\qquad E^-(z)E^-(w)=reg.,\nonumber \\
E^+(z)E^-(w)&=&{k\over(z-w)^2}+ {\sqrt 2 J(w)\over{z-w}}+reg.,\nonumber
\\ J(z)J(w)&=&{k\over(z-w)^2}+reg.  \eea

The $su(2)/u(1)$  Virasoro generator $K(w)$ is given by
\bea  \label {su(2)}
K(w)=\sum_m {K(m)w^{-m-2}}= L^{su(2)}(w)-L^{u(1)}(w),
\eea
where
\bean
L^{su(2)}(w)&=&\frac 1 {2(k+2)}\left (:J(w)J(w):+:E^+(w)E^-(w):+
:E^-(w)E^+(w):\right ),  \nonumber \\
L^{u(1)}(w)&=&\frac 1 {2k}:J(w)J(w):.
\eean
 The current $J(w)=\sum_m {J(m)w^{-m-1}}$ has unit
conformal dimension with respect to
$L^{su(2)}(w)=\sum_m {L^{su(2)}(m)w^{-m-2}}
$ and  $L^{u(1)}(w)=\sum_m {L^{u(1)}(m)w^{-m-2}}$
\bean
[L^{su(2)}(m),J(n)]=[L^{u(1)}(m),J(n)]=-nJ^i(m+n).
\eean

According to eq. (\ref {sol}) the coset currents are given by
\bean
\tilde E^+(w)&=&e^{-\frac {\sqrt 2} k  Q_<(w)}
E^+(w)e^{-\frac {\sqrt 2} k  Q_>(w)}
w^{-\frac {\sqrt 2}  k J(0)},
\nonumber \\
\tilde E^-(w)&=&e^{\frac {\sqrt 2} k  Q_<(w)}
E^-(w)e^{\frac {\sqrt 2} k Q_>(w)}
w^{\frac {\sqrt 2}  k J(0)}                 .
\eean
These fields are closely related with $Z$ operators for
$su(2)$  \cite {LW,LP}.

Using eqs. (\ref {OPE}-\ref {su(2)}) one can check that
the currents
\bea \label{par}
\psi_1^{\phantom {l}} (w)=(1/\sqrt k) \tilde E^+(w),\qquad
\psi^+_1(w)=(1/\sqrt k) \tilde E^-(w) \eea satisfy the parafermion
algebra of Ref. \cite {FZ}
\bean\label{}
\psi_1^{\phantom {l}}(z)\psi_1^+(w)=(z-w)^{-2+{2\over
k}}\left(I+{{k+2}\over k}K(w)(z-w)^2+O\left((z-w)^3\right)\right),
\eean
where $K(w)$ is given by (\ref {su(2)}).Eqs. (\ref
{par}) give a new realization of the parafermion currents
$\psi_1^{\phantom {l}}(w)$ and $\psi_1^+(w).$
\bigskip

I would like to thank J.Lepowsky for useful correspondence.


\end{document}